\documentclass[sigconf,nonacm]{acmart}

\usepackage{graphicx}
\usepackage{subfiles}

\usepackage{todonotes}
\usepackage{booktabs}
\PassOptionsToPackage{hyphens}{url}\usepackage{hyperref}

\usepackage{dsfont}
\usepackage{amsmath}

\usepackage{caption}
\usepackage{subcaption}

\usepackage{lipsum}  

\usepackage{xcolor}

\usepackage{pifont}

\AtBeginDocument{%
  \providecommand\BibTeX{{%
    \normalfont B\kern-0.5em{\scshape i\kern-0.25em b}\kern-0.8em\TeX}}}

\begin{document}

\title{HLTCOE at TREC 2024 NeuCLIR Track}

\author{Eugene Yang, Dawn Lawrie, Orion Weller, James Mayfield}
\affiliation{%
  \institution{Human Language Technology Center of Excellence, Johns Hopkins University, USA}
  \country{}}
\email{{eugene.yang,lawrie,oweller2,mayfield}@jhu.edu}

\renewcommand{\shortauthors}{Yang et al.}

\begin{abstract}
The HLTCOE team applied PLAID, an mT5 reranker, GPT-4 reranker, score fusion,
and document translation to the TREC 2024 NeuCLIR track.
For PLAID we included a variety of models and training techniques --
Translate Distill~(TD), Generate Distill~(GD) and multilingual translate-distill~(MTD). 
TD uses scores from the mT5 model over English MS~MARCO query-document pairs
to learn how to score query-document pairs where the documents are translated to match the CLIR setting.
GD follows TD but uses passages from the collection and queries generated by an LLM for training examples.
MTD uses MS~MARCO translated into multiple languages,
allowing experiments on how to batch the data during training.
Finally, for report generation we experimented with system combination over different runs.
One family of systems used either GPT-4o or Claude-3.5-Sonnet
to summarize the retrieved results from a series of decomposed sub-questions.
Another system took the output from those two models and verified/combined them with Claude-3.5-Sonnet.
The other family used GPT4o and GPT3.5Turbo
to extract and group relevant facts from the retrieved documents based on the decomposed queries.
The resulting submissions directly concatenate the grouped facts to form the report
and their documents of origin as the citations. 
The team submitted runs to all NeuCLIR tasks:
CLIR and MLIR news tasks as well as the technical documents task and the report generation task.
\end{abstract}


\maketitle

\section{Introduction}

This year the HLTCOE's primary contribution was in experimentation with multiple ways to fine-tune mPLMs for CLIR.
Specifically, we submitted a suite of ColBERT-X models,
including Translate-Distill~\cite{tdistill} and Generate-Distill (GD),
Each model relied on an effective cross-encoder developed for the NeuCLIR 2022 track as the teacher~\cite{unicamp-at-neuclir},
applying scores to English queries and documents.
We compared Translate-Distill in the CLIR setting against the zero-shot setting
where the model sees English queries and documents rather than English queries and non-English documents.
The HLTCOE also experimented with Generate-Distill where the ColBERT-X model is fine-tuned 
using queries generated by an LLM, following the Translate-Distill approach for CLIR.
A few of our submissions investigated the effect of reranked runs with the mT5 cross-encoder 
and GPT-4 with heapsort. 
Finally, we implemented token pooling~\cite{clavié2024reducingfootprintmultivectorretrieval} in PLAID,
reducing the number of tokens in the index by half.
Table~\ref{tab:run-name_clir} summarizes our CLIR runs in the news domain; 
Table~\ref{tab:run-name_mlir} summarizes our MLIR runs; and 
Table~\ref{tab:run-name_tech} summarizes our technical documents runs.\footnote{Given that the authors are also organizers of the track, all hltcoe runs are marked as manual runs. Although unlikely, performance might have been affected by knowledge that was only accessible to the organizers.}

The HLTCOE also participated in the pilot report generation task.
We produced two families of systems.
The first proposed a summarization-based approach
to decompose the original report topic into a variable length number of subquestions.
Those subquestions then were used to find relevant documents
through the search service provided by the track coordinators.
Then all subquestions and results were presented to the language model (either Claude-3.5-Sonnet or GPT-4o).
A meta-system combined the results from both GPT-4o and Claude-3.5-Sonnet with the retrieved documents,
fact checking and updating both of them.

The other family extracted relevant facts from the retrieved documents. 
This family also decomposed the topic into multiple queries
and use the search service to retrieve candidate documents.
The facts extracted from the documents are grouped into canonical facts.
The canonical fact descriptions are directly used as the report sentence
and the documents that contributed to each group are used as the citations. 

In the rest of this paper, we describe our systems, submissions, and observations.

\begin{table*}[]
\caption{HLTCOE CLIR runs over newswire collection. XXX indicates the language of the newswire collection. }\label{tab:run-name_clir}
    \centering

\resizebox{\linewidth}{!}{
\begin{tabular}{cl|cl|c|l}
\toprule
& Run Name & Type &  Model & Query & Description \\
\midrule
(c1) & plaid\_eqsynms\_distill\_engXXX                   & Dense &          PLAID & TD & GD/TD with an equal number of generate  \\
 & & & & & \hspace{2em}queries and MS~MARCO queries \\
 (c2) & kitchen\_rankfuse.mt5rerank.gpt4rerank                & Hybrid & multiple >> mT5 >> GPT & TD & Rerank top 30 c4 run with gpt4 using  \\
 & & & & & \hspace{2em}heapsort \\
 (c3) & plaid\_distill\_engXXX.mt5rerank                &  Hybrid &  PLAID >> mT5 & TD & Rerank c6 with mt5 \\
 (c4) & kitchen\_rankfuse.mt5rerank                      &  Hybrid & multiple >> mT5    & TD & Rerank c9 with mt5 \\ 
 (c5) & plaid\_syn\_distill\_engXXX                       &  Dense &        PLAID & TD & Trained with GD \\ 
 (c6) & plaid\_distill\_engXXX                  &  Dense &        PLAID & TD & TD on English query and translated  \\
 & & & & & \hspace{2em} MS~MARCO documents \\ 
 (c7) & plaid\_distill\_engeng\_zs2engXXX                           & Dense &        PLAID &  TD & Distill on English MS~MARCO, zero shot to \\
 & & & & & \hspace{2em}English qry and native docs \\
(c8) & plaid\_distill\_engXXX.mt5rerank.gpt4rerank          & Hybrid & PLAID >> mt5 >> GPT & TD & Rerank top 30 t3 run with gpt4 using   \\
 & & & & & \hspace{2em}heapsort \\
(c9) & kitchen\_rankfuse                           & Hybrid &        multiple &  TD & ranked-based fusion over c5, c10, \\
& & & & & \hspace{2em}coord...-MNES-fast\_psq\_td, \\
& & & & & \hspace{2em}coord...-MNES-patapscoBM25qtRM3td, \\
& & & & & \hspace{2em}coord...-MTES-patapscoBM25dtRM3td, \\
& & & & & \hspace{2em}and ISI\_SEARCHER-ANE\_run1 \\
(c10) & plaid\_distill\_engeng                   &  Dense &        PLAID & TD & Distill on English MS~MARCO, indexing \\
& & & & & \hspace{2em}translated documents \\ 
(c11) & plaid\_distill\_mono\_XXXXXX       &Dense & PLAID  & TD & TD on translated MS~MARCO queries and \\
& & & & & \hspace{2em}documents, run with GT queries \\
(c12) & plaid\_distill\_engeng\_zs2XXXXXX             & Dense &      PLAID & TD & Distill on English MS~MARCO, zero shot to \\
& & & & & \hspace{2em}GT queries \\
(c13) & plaid\_distill\_engXXX\_450p             & Dense &      PLAID & TD & TD on English query and translated   \\
 & & & & & \hspace{2em}MS~MARCO documents, indexing  \\
& & & & & \hspace{2em}passages of 450 tokens without overlap \\
(c14) & plaid\_distill\_engmlir             & Dense &      PLAID & TD & TD on English query and translated \\
& & & & & \hspace{2em}MS~MARCO into all 3 NeuCLIR  \\
& & & & & \hspace{2em}languages \\
(c15) & plaid\_distill\_engXXX\_termpool2             & Dense &      PLAID & TD & TD on English query and translated   \\
 & & & & & \hspace{2em}MS~MARCO documents, indexed with \\
 & & & & & \hspace{2em}term pooling removing $1/2$ tokens \\ 
\bottomrule
\end{tabular}
}

\end{table*}

\begin{table*}[]
\caption{HLTCOE MLIR runs over newswire collection. }\label{tab:run-name_mlir}
    \centering

\resizebox{\linewidth}{!}{
\begin{tabular}{cl|cl|c|l}
\toprule
& Run Name & Type &  Model & Query & Description \\
\midrule
(m1) & plaid\_distill\_mlir\_bycoll\_scorefuse                             &  Dense &    PLAID & TD & MTD with mixed entries indexing each  \\
& & & & & \hspace{2em}language separately; fusing results \\
& & & & & \hspace{2em}by scores \\
(m2) & plaid\_distill\_clir.mt5rerank.scorefuse.gpt4rerank                 &  Dense &        PLAID >> mT5 >> GPT & TD & Reranked top 30 of run m10 with GPT-4 \\
(m3) & kitchen\_rankfuse.mt5rerank.scorefuse.gpt4rerank                    &  Dense &        multiple >> mT5 >> GPT & TD & Reranked top 30 of run mX with GPT-4 \\
(m4) & plaid\_distill\_engeng           &  Dense &        PLAID & TD & PLAID trained on English MS~MARCO with \\
& & & & & \hspace{2em}a single index of translated documents \\
(m5) & plaid\_distill\_engeng\_zs           &  Dense &        PLAID & TD & PLAID trained on English MS~MARCO with \\
& & & & & \hspace{2em}a single index of native documents \\
(m6) & plaid\_distill\_mlir\_mixedentry           &  Dense &        PLAID & TD & MTD PLAID trained with mixed entries \\
& & & & & \hspace{2em}using a single index of native documents \\
(m7) & plaid\_distill\_mlir\_mixedpass           &  Dense &        PLAID & TD & MTD PLAID trained with mixed passages \\
& & & & & \hspace{2em}using a single index of native documents \\
(m8) & plaid\_distill\_mlir\_rr           &  Dense &        PLAID & TD & MTD PLAID trained with round robin using \\
& & & & & \hspace{2em}a single index of native documents \\
(m9) & plaid\_distill\_mlir\_mixedentry\_termpool2           &  Dense &        PLAID & TD & MTD PLAID trained with mixed entries \\
& & & & & \hspace{2em}using a single index of native documents \\
& & & & & \hspace{2em}indexed with term pooling removing \\
& & & & & \hspace{2em}$1/2$ token \\
(m10) & plaid\_distill\_clir.mt5rerank.scorefuse           &  Dense &        PLAID >> mT5 & TD & Reranked run m11 with mT5 \\
(m11) & plaid\_distill\_clir\_scorefuse           &  Dense &        PLAID & TD & Fused scores of CLIR runs c5 for each \\
& & & & & \hspace{2em}language \\
(m12) & kitchen\_rankfuse.mt5rerank.scorefuse           &  Dense &        multiple >> mT5 & TD & Fused scores of CLIR runs c4 for each \\
& & & & & \hspace{2em}language \\

\bottomrule
\end{tabular}
}

\end{table*}

\begin{table*}[]
\caption{HLTCOE runs over technical documents. }\label{tab:run-name_tech}
    \centering

\resizebox{\linewidth}{!}{
\begin{tabular}{cl|cl|c|l}
\toprule
& Run Name & Type &  Model & Query & Description \\
\midrule

 (t1) & plaid\_eqsynms\_distill\_engzho                   & Dense &          PLAID & TD & GD/TD with an equal number of generate queries \\
 & & & & & \hspace{2em}and MS~MARCO queries \\
 (t2) & kitchen\_rankfuse.mt5rerank.gpt4rerank                & Hybrid & multiple >> mT5 >> GPT & TD & Rerank top 30 t4 run with gpt4 all at once \\
 (t3) & plaid\_distill\_engzho.mt5rerank                &  Hybrid &  PLAID >> mT5 & TD & Rerank t6 with mt5 \\
 (t4) & kitchen\_rankfuse.mt5rerank                      &  Hybrid & multiple >> mT5    & TD & Rerank t9 with mt5 \\ 
 (t5) & plaid\_syn\_distill\_engzho                       &  Dense &        PLAID & TD & Trained with GD \\ 
 (t6) & plaid\_distill\_engzho                  &  Dense &        PLAID & TD & TD on English query and translated MS~MARCO  \\
 & & & & & \hspace{2em} documents \\ 
 (t7) & plaid\_distill\_engeng\_zs2engzho                           & Dense &        PLAID &  TD & Distill on English MS~MARCO, zero shot to Eng. qry \\
 & & & & & \hspace{2em}and native docs \\
(t8) & plaid\_distill\_engzho.mt5rerank.gpt4rerank          & Hybrid & PLAID >> mt5 >> GPT & TD & Rerank top 30 t3 run with gpt4 all at once  \\
(t9) & kitchen\_rankfuse                           & Hybrid &        multiple &  TD & ranked-based fusion over t5, t10, \\
& & & & & \hspace{2em}coordinators-MNES-fast\_psq\_td, \\
& & & & & \hspace{2em}coordinators-MNES-patapscoBM25qtRM3td, \\
& & & & & \hspace{2em}coordinators-MTES-patapscoBM25dtRM3td, \\
& & & & & \hspace{2em}and ISI\_SEARCHER-ANE\_run1 \\
(t10) & plaid\_distill\_engeng                   &  Dense &        PLAID & TD & TD on English MS~MARCO, indexing \\
& & & & & \hspace{2em}translated documents \\ 
(t11) & plaid\_distill\_zhozho       &Dense & PLAID  & TD & TD on translated MS~MARCO queries and \\
& & & & & \hspace{2em}documents, run with GT queries \\
(t12) & plaid\_distill\_engeng\_zs2zhozho             & Dense &      PLAID & TD & Distill on English MS~MARCO, zero shot to GT \\
& & & & & \hspace{2em}queries \\
\bottomrule
\end{tabular}
}

\end{table*}

\section{mT5 Reranker}

At NeuCLIR 2022, the most effective submission was a reranking model
that used the MonoT5 model with mT5XXL, a 13 billion parameter model~\cite{neuclir2022, unicamp-at-neuclir}. 
This submission uses pointwise reranking
by taking a softmax over the decoding probabilities of two specific tokens
(\texttt{\_false} and \texttt{\_true} for mT5XXL)
as the probability of the document being relevant given the query~\cite{monot5}. 
In this paper, we use the name ``mT5 reranker'' to refer to this specific reranker for convenience. 
The input queries are titles concatenated with descriptions.

\section{GPT-4 Reranker}

We used GPT-4o in two ways for our submissions.
For the Technical documents track,
we reranked the 30 top ranked documents from the run, using the same prompt used by \citet{Parry2024}.
For the news collection, we first had GPT-4 determine the most relevant passage
in each of the top 30 documents in the run being used for reranking.
Then we performed 4-ary heapsort~\cite{heapsort25} on these top 30 passages.
Each passage contained 450 tokens.
The prompt we used from this task was slightly modified from the one used by \citet{Parry2024}:
\begin{quote}
SYSTEM\_MESSAGE
You are an intelligent assistant that can identify the best passage based on its relevance to a query.

\vspace{1em}

PROMPT

I will provide you with 5 passages in no particular order, each indicated by a numerical identifier in square brackets.
      Identify the best passage based on its relevance to this search query: \{description\}.

[1] ...

[2] ...

[3] ...

[4] ...

[5] ...

General search topic: \{title\} Search Query: \{description\}.

Identify the best passage above based on its relevance to the search query.
      The output format should be [], e.g., [4] meaning document 4 is most relevant to the search query.
      Only respond with the passage number; do not say any word or explain.

\end{quote}

If during passage selection or a heapsort call GPT4 failed to return a document id,
the first document was selected.
For passage selection, this seemed reasonable
because generally the main information in a news article is summarized in the beginning of the article.
For heapsort this default may not always be optimal.
For initial heap construction the decision is likely correct
because the documents had already been sorted by the prior algorithm.
However, this is less true when using heapify to rebuild the heap
because heapsort is not a stable sorting algorithm.
Heapsort was used to identify the top 20 passages.
The remaining passages were left in an unsorted order since the primary measure used in NeuCLIR is nDCG@20. 

\section{ColBERT Retrieval with PLAID}
Our systems primarily use PLAID~\cite{plaid},
an implementation of the ColBERT~\cite{colbert} retrieval architecture
that encodes each token as a vector.
Prior work on training CLIR dense retrieval models has demonstrated success
augmenting training queries and passages with translation to match the CLIR target languages~\cite{colbertx}.
In the NeuCLIR 2023 track~\cite{neuclir2023},
our ColBERT-X models trained with Translate-Distill
were the most effective end-to-end neural dense retrieval models,
so that was the starting point for this year's submissions.

All ColBERT-X variants use the PLAID-X implementation with the title concatenated with the description as the query. 
Unless otherwise noted, we break each document into passages of 180 tokens with a stride of 90. 
We use one bit for each dimension of the residual vector.
We search for 2500 passages and aggregate passage scores into a document scores using MaxP~\cite{maxp}.

\subsection{Translate-Distill for CLIR using mT5 Reranker}
Distillation is an effective strategy for training a small yet efficient model
that mimics the effectiveness of a larger and computationally more expensive model~\cite{formal2021splade, rocketqa}. 
In our submissions, we explored training a ColBERT-X model to mimic the behavior of the powerful mT5 reranker. 
This training method, known as Translate-Distill,  is described in more detail in \citet{tdistill}.

Translate-Distill starts by selecting hard passages for each training query in the MS~MARCO training set. 
We use an \textbf{English} ColBERTv2 model~\cite{colbertv2} to retrieve the top 50 passages for each query
from the MS~MARCO training set.
To obtain the teacher scores for each query/passage pair,
the mT5 reranker scores the \textbf{translated} passage
along with the English query.\footnote{\citet{tdistill} finds that this configuration is not the optimal way to obtain scores. Instead English query/passage pairs should be used to obtain scores.}
Finally, we train the ColBERT-X model using these hard passages in the document language,
queries in English, and scores from the mT5-xxl reranker with a KL Divergence loss. 
Translate-Distill is an extension of the ColBERT-X Translate-Train~\cite{colbertx} approach
that distills ranking knowledge from a reranker instead of learning from the contrastive labels. 

\subsection{Generate-Distill for CLIR using mT5 Reranker}

Generate-Distill~\cite{GD25} is aimed at overcoming domain mismatch between the training and search data
and the training bottleneck caused by translation quality.
This approach fine-tunes a CLIR retrieval model on the search corpus directly. 
Since we assume no queries or relevance judgments are available at training time, we use the model \texttt{Mixtral-8x7B-Instruct}\footnote{\url{https://huggingface.co/mistralai/Mixtral-8x7B-Instruct-v0.1}} to generate training queries
and the mt5 cross-encoders as a teacher model to provide their associated relevance signals for CLIR retrieval fine-tuning.
Once the queries are generated, we follow the same process as Translate-Distill to train the model.

\subsection{Token Pooling to Reduce Index Size}

We submitted runs that integrated the technique of token pooling 
introduced by \citet{clavié2024reducingfootprintmultivectorretrieval} with PLAID-X.
This technique pools a passage's tokens
and uses k-means clustering to determine the representative vectors for the tokens.
We submitted runs that reduced the number of tokens by half.

\begin{table*}[t]
\caption{CLIR Results}\label{tab:clir-results}
    \centering

\begin{tabular}{l|ccc|ccc}
\toprule
{} & \multicolumn{3}{c}{nDCG@20} & \multicolumn{3}{c}{R@1000} \\

& Persian & Russian & Chinese & Persian & Russian & Chinese \\

\midrule
kitchen\_rankfuse.mt5rerank.gpt4rerank     &   0.690 &  0.585 &  0.664 &  0.974 &  0.984 &  0.993 \\
plaid\_distill\_engXXX.mt5rerank.gpt4rerank &   0.676 &  0.592 &  0.650 &  0.917 &  0.968 &  0.947 \\
kitchen\_rankfuse.mt5rerank                &   0.619 &  0.509 &  0.584 &  0.974 &  0.984 &  0.993 \\
plaid\_distill\_engXXX.mt5rerank            &   0.610 &  0.508 &  0.582 &  0.917 &  0.968 &  0.947 \\
kitchen\_rankfuse                          &   0.629 &  0.531 &  0.569 &  0.974 &  0.984 &  0.993 \\
plaid\_distill\_engXXX\_450p                 &   0.611 &  0.543 &  0.562 &  0.928 &  0.967 &  0.941 \\
plaid\_distill\_engmlir                     &   0.592 &  0.497 &  0.560 &  0.923 &  0.961 &  0.955 \\
plaid\_distill\_engXXX\_termpool2            &   0.600 &  0.504 &  0.555 &  0.916 &  0.963 &  0.941 \\
plaid\_distill\_engXXX                      &   0.607 &  0.508 &  0.552 &  0.917 &  0.968 &  0.947 \\
plaid\_distill\_engeng                      &   0.605 &  0.498 &  0.541 &  0.940 &  0.962 &  0.961 \\
plaid\_syn\_distill\_engXXX                  &   0.578 &  0.490 &  0.540 &  0.915 &  0.954 &  0.954 \\
plaid\_distill\_engeng\_zs2XXXXXX            &   0.587 &  0.478 &  0.527 &  0.928 &  0.947 &  0.940 \\
plaid\_distill\_mono\_XXXXXX                 &   0.574 &  0.493 &  0.520 &  0.912 &  0.952 &  0.929 \\
plaid\_distill\_engeng\_zs2engXXX            &   0.541 &  0.514 &  0.499 &  0.909 &  0.950 &  0.920 \\
plaid\_eqsynms\_distill\_engXXX              &   0.622 &  0.504 &  0.490 &  0.940 &  0.960 &  0.922 \\
\bottomrule
\end{tabular}

\end{table*}

\subsection{Multilingual Translate-Distill for MLIR using mT5 Reranker}

We submitted the multilingual ColBERT-X models described in \citet{yang2024distillation},
which is an extended version of the Translate-Distill for CLIR~\cite{tdistill}.
We submitted all three variants of the Multilingual Translate-Distill (MTD),
including mixed entry (\texttt{mixedentry}), mixed passages (\texttt{misedpass}), and round robin (\texttt{rr}). 
Each model is initialized with XLM-RoBERTa-Large and trained with translated MS MSMARCO provided by the coordinators~\cite{neuclir2022}. 
The teacher scores are obtained by applying the mT5-xxl cross-encoder on the selected query-passage pairs,
which are the same as those used in the CLIR version. 

\section{Report Generation Systems}
We developed two families of systems: bottom-up and top-down.

\paragraph{Summarization-based Approach}
The summarization-based approach used a language model with the initial topic as input,
generated a persona for that topic,
split the topic into a variable number of subqueries to find information,
then aggregated that information using a language model.
The prompt was tuned to help the model have an unbiased tone, cite properly, and not cite too little or too much. GPT-Researcher\footnote{\url{https://github.com/assafelovic/gpt-researcher}} was used as the framework for this approach.

We used two language models, GPT-4o-08-06 and Claude-3.5-Sonnet.
These two models were chosen because of their strong performance in the NeuCLIR languages.
The language model received all subqueries and the top ten retrieved documents for each subquery.
It then summarized these elements into a report. 

Finally, as these models can be prone to hallucination,
we developed a meta-system that took in both reports as well as the subquestions and retrieved documents,
and created a final report.
This allowed for self-reflection and updating to ensure that the report correctly found all aspects in the retrieved documents.

The submission IDs of this approach are \texttt{\{lang\}-jhu-orion-agg\\regated-w-\{gpt4o,claude\}}.

\paragraph{Extractive Approach. }
The extractive approach also uses a language model to generate the final report.
Despite conceptually being extractive,
the pipeline still heavily uses the language model to ``extract'' information
instead of using a traditional BIO-style extraction model.
In this approach, we use the framework DSPy\cite{khattab2024dspy,khattab2022demonstrate} and the rely on the models GPT-4o-08-06 and GPT3.5Turbo. 

Similar to the summarization-based approach,
the extractive approach also first splits the topic into multiple queries.
Specifically, the generative model is prompted to generate ten ``Google-like'' queries based on the report request.
The queries are then issued to the search service provided by the coordinators to obtain the top three documents for each query. 

For each retrieved documents, we prompted the generative model to extract key facts based on the report requests. 
The model then groups the extracted facts into twenty ``canonical'' facts from all retrieved documents
with the document ID being attached to each group. 

Since the evaluation only rewards individual sentences in scoring reports
instead of favoring a coherent, human-readable report,
we treat each canonical fact as the report sentence and the documents containing the fact as the citations. 

The submission IDs of this approach are \texttt{\{lang\}-hltcoe-eugene-\\\{gpt4o,gpt35turbo\}}.

\section{Results}

\subsection{CLIR Tasks}

The CLIR runs submitted by the HLTCOE are summarized in Table~\ref{tab:clir-results}.
The most effective run among our submissions is reranking using 4-ary heapsort.
For Chinese and Persian, c2 was the most effective using a large variety of other runs.
For Russian starting with TD PLAID in run c8 was the most effective.
Systems in the second tier of algorithm performance were much less uniform by language;
however, using passages of 450 tokens was always among them (c13). 
Interestingly, only for Chinese was reranking with mt5 more effective than just fusing many disparate runs (c4).
Finally the introduction of Generate-Distill (c1) was most effective for Persian. 

\subsection{MLIR Task}
\begin{table}[t]
\caption{MLIR Runs}\label{tab:mlir-results}
    \centering

\resizebox{\linewidth}{!}{
\begin{tabular}{l|cc }
\toprule
{} &  nDCG@20 &  R@1000 \\
\midrule
plaid\_distill\_clir.mt5rerank.scorefuse.gpt4rerank &   0.460 &  0.876 \\
kitchen\_rankfuse.mt5rerank.scorefuse.gpt4rerank   &   0.459 &  0.899 \\
plaid\_distill\_clir.mt5rerank.scorefuse            &   0.405 &  0.876 \\
kitchen\_rankfuse.mt5rerank.scorefuse              &   0.404 &  0.899 \\
plaid\_distill\_mlir\_mixedpass                      &   0.383 &  0.877 \\
plaid\_distill\_mlir\_mixedentry\_termpool2           &   0.377 &  0.871 \\
plaid\_distill\_mlir\_mixedentry                     &   0.376 &  0.874 \\
plaid\_distill\_engeng                              &   0.373 &  0.875 \\
plaid\_distill\_mlir\_bycoll\_scorefuse               &   0.365 &  0.875 \\
plaid\_distill\_mlir\_rr                             &   0.360 &  0.862 \\
plaid\_distill\_engeng\_zs                           &   0.297 &  0.827 \\
plaid\_distill\_clir\_scorefuse                      &   0.296 &  0.799 \\
\bottomrule
\end{tabular}
}

\end{table}

The results of our MLIR submissions are summarized in Table~\ref{tab:mlir-results}.
The most effective run among our submissions is m2,
which relies on GPT-4 to execute the heapsort algorithm over the results of PLAID TD.
Just below the GPT-4 reranking is the mT5 re-ranking,
which may indicate that for MLIR to cope with documents in multiple languages
it is advantageous to view the query and document at the same time.

\subsection{Technical Documents Task}

\begin{table}[t]
\caption{Technical Document Track Results. Italicized names indicate monolingual runs.}\label{tab:tech-results}
    \centering

\resizebox{\linewidth}{!}{
\begin{tabular}{l|cc}
\toprule
{} &  nDCG@20 &  R@1000 \\
\midrule

plaid\_distill\_engzho.mt5rerank.gpt4rerank &   0.455 &  0.838 \\
kitchen\_rankfuse.mt5rerank.gpt4rerank     &   0.454 &  0.919 \\
kitchen\_rankfuse.mt5rerank                &   0.434 &  0.919 \\
plaid\_distill\_engzho.mt5rerank            &   0.429 &  0.838 \\
plaid\_eqsynms\_distill\_engzho              &   0.385 &  0.867 \\
plaid\_distill\_engeng\_zs2zhozho            &   0.379 &  0.851 \\
kitchen\_rankfuse                          &   0.378 &  0.919 \\
plaid\_distill\_zhozho                      &   0.372 &  0.846 \\
plaid\_distill\_engeng                      &   0.370 &  0.834 \\
plaid\_distill\_engzho                      &   0.368 &  0.838 \\
plaid\_syn\_distill\_engzho                  &   0.356 &  0.838 \\
plaid\_distill\_engeng\_zs2engzho            &   0.265 &  0.736 \\

\bottomrule
\end{tabular}
}
\end{table}

The technical documents task represents a huge domain shift from news documents and from the MS~MARCO training data,
which in turn leads to different algorithm rankings.
The HLTCOE submitted many of the same variants to this task as it did for the news tasks.
Table~\ref{tab:tech-results} shows that GPT4 is more effective than the other techniques.
After reranking, combining TD and GD is the most effective of the dense retrieval approaches. 

\subsection{Report Generation Task}

\begin{table*}[]
\caption{ Report Generation Task Scores.  }\label{tab:repgen-results}
\centering

\begin{tabular}{ll|c|cccc}
\toprule
{} & {}                                               & ARGUE         & Citation      & Nugget        & Nugget        & Sentence      \\
Language & Run ID                                           & Score         & Precision     & Recall        & Support       & Support       \\
\midrule

Chinese& zho-hltcoe-eugene-gpt4o-fixed                    & 0.726  & 0.859  & 0.327  & 0.298  & 0.840  \\
& zho-hltcoe-eugene-gpt35turbo                     & 0.587  & 0.813  & 0.250  & 0.248 & 0.720  \\
& zho-jhu-orion-aggregated-w-claude                & 0.528  & 0.900  & 0.177  & 0.238  & 0.662  \\
&zho-jhu-orion-aggregated-w-gpt4o                 & 0.496  & 0.899  & 0.182  & 0.237  & 0.636  \\

\midrule
Persian & fas-hltcoe-eugene-gpt4o                          & 0.872  & 0.918  & 0.303  & 0.311  & 0.919  \\
 & fas-hltcoe-eugene-gpt35turbo                     & 0.646  & 0.846 & 0.215  & 0.201  & 0.792  \\
 &fas-jhu-orion-aggregated-w-claude                & 0.519  & 0.945  & 0.213  & 0.268  & 0.650  \\
 & fas-jhu-orion-aggregated-w-gpt4o                 & 0.449  & 0.941  & 0.220  & 0.272  & 0.565  \\

\midrule
Russian & rus-hltcoe-eugene-gpt4o                          & 0.808  & 0.904  & 0.339  & 0.420  & 0.874  \\
 &  rus-hltcoe-eugene-gpt35turbo                     & 0.602  & 0.799  & 0.313  & 0.309  & 0.715  \\
 & rus-jhu-orion-aggregated-w-claude                & 0.519  & 0.902  & 0.255  & 0.323  & 0.673  \\
 &  rus-jhu-orion-aggregated-w-gpt4o                 & 0.415  & 0.904  & 0.247  & 0.287  & 0.495  \\

\bottomrule
\end{tabular}

\end{table*}

The HLTCOE submitted four runs per language to the report generation task. Table~\ref{tab:repgen-results} contains the results. While the report request was the same for each language, the documents over which the report was generated varied by language. Overall, system ranking is consistent by language. Our extractive approach performed better than our abstractive approach in terms of the ARGUE score and  nugget recall. This is not surprising, as the extractive approach was designed to favor the inclusion of more facts over producing a fluent report. 
Given the underlying LLMs used to assist the report creation process, GPT3.5 struggled with the task more than the other models, especially in terms of citation precision.

\section{Conclusion}

The HLTCOE team participated in all tasks offered in NeuCLIR 2024.
Officially all runs were marked as manual because, as organizers,
we had input into creating the document collections,
running the annotation for the technical documents task,
and writing the guidelines for the tasks.
In general, runs reranked with GPT4 outperformed runs reranked with mT5,
which in turn outperformed end-to-end neural approaches
(although this general trend did not hold for CLIR in the news domain where mT5 ranking was not as helpful).
Increasing the passage size to 450 tokens and not overlapping the passages led to a clear performance boost,
while decreasing the size of the index by half.
Reducing the index size using token pooling was less effective and slowed indexing time substantially. 
For report generation, optimizing the report to include as many facts as possible is beneficial.
Directions for future research include more investigation into training CLIR and MLIR models using generate-distill.

\bibliographystyle{ACM-Reference-Format}
\bibliography{sample-base}

\end{document}